\documentclass[aps,prb,twocolumn,superscriptaddress,showpacs,english]{revtex4-1}

\usepackage[T1]{fontenc}
\usepackage{lipsum}

\usepackage[latin9]{inputenc}
\usepackage{babel}
\usepackage{amsmath}
\usepackage{amssymb}
\usepackage{bbold}
\usepackage{wasysym}
\usepackage{graphicx}
\usepackage{multirow}
\usepackage{gensymb}
\usepackage[dvipsnames]{xcolor}
\usepackage{pifont}
\usepackage{microtype}
\usepackage{physics}
\usepackage{xcolor}
\usepackage{soul}
\usepackage[version=3]{mhchem}
\usepackage{chemformula}[2014/04/08] 
\usepackage{booktabs}    
\usepackage{dcolumn}     
\usepackage{bm}          
\usepackage{endnotes}

\usepackage[linktocpage=true,
  colorlinks=true, 
  pdfborder={0 0 0},
  linkcolor=blue,
  citecolor=red,
  filecolor=yellow,
  urlcolor=blue,
  bookmarks,
  pdfauthor={},
]{hyperref}

\usepackage{longtable}

\newcommand{\Unidonostia}{Fisika Aplikatua Saila, Gipuzkoako Ingeniaritza Eskola, University of the Basque Country (UPV/EHU), Europa Plaza 1, 20018 Donostia/San Sebasti\'an, Spain}
\newcommand{\CFM}{Centro de F\'isica de Materiales (CSIC-UPV/EHU), Manuel de Lardizabal Pasealekua 5, 20018 Donostia/San Sebasti\'an, Spain}
\newcommand{\DIPC}{Donostia International Physics Center (DIPC),  Manuel de Lardizabal Pasealekua 4, 20018 Donostia/San Sebasti\'an, Spain}

\DeclareMathAlphabet\mathbfcal{OMS}{cmsy}{b}{n}

\begin{document}

\title{Impact of ionic quantum fluctuations on the thermodynamic stability and superconductivity of LaBH$_{8}$}

\author{Francesco Belli}
\affiliation{\CFM} \affiliation{\Unidonostia}

\author{Ion Errea}
\affiliation{\CFM} \affiliation{\Unidonostia} \affiliation{\DIPC}

\begin{abstract}
The recent prediction of a metastable high-symmetry $Fm\bar{3}m$ phase of LaBH$_8$ gives hopes to reach high superconducting critical temperatures at affordable pressures among ternary hydrogen-rich compounds. Making use of first-principles calculations within density functional theory and the stochastic self-consistent harmonic approximation, we determine that ionic quantum fluctuations drive the system dynamically unstable below 77 GPa, a much higher pressure than the 45 GPa expected classically. Quantum anharmonic effects stretch the covalent B-H bond in the BH$_8$ units of the structure and, consequently, soften all hydrogen-character modes. Above 77 GPa $Fm\bar{3}m$  LaBH$_8$ remains metastable and, interestingly, its superconducting critical temperature is largely enhanced by quantum anharmonic effects, reaching critical temperatures around 170 K at the verge of the dynamical instability. Our results suggest that low pressure metastable phases with covalently bonded symmetric XH$_8$ units will be destabilized by ionic quantum fluctuations. 
\end{abstract}

\maketitle



\section{Introduction}

The quest for room-temperature superconductors is one of the most long-awaited goals in condensed matter physics. Hydrogen-rich systems are among the most promising candidates as, originally suggested by Neil Ashcroft \cite{Ashcroft,Ashcroft2}, their large Debye temperature allows in principle the possibility to reach a high superconducting critical temperature ($T_c$). In the last few years, the discovery of H$_3$S \cite{2}, with a superconducting critical temperature of 200K at 150 GPa, confirmed that hydrogen-rich materials can reach $T_c$ values approaching room temperature. This result was followed by further experiments that measured critical temperatures around 250 K at megabar pressures in YH$_6$ \cite{3}, LaH$_{10}$ \cite{4,5}, and YH$_9$ \cite{6,Snider2021Synthesis}. Finally, room temperature superconductivity has been claimed in a compound composed by a mixture of hydrogen, carbon, and sulfur, with a critical temperature of 288 K at about 270 GPa \cite{7}. On the downside, however, much work is still needed to reduce the pressure of stability of these high-$T_c$ hydrides, in the order of megabars, impractical for any technological use of this class of compounds.

The potential of first-principles calculations based on density-functional theory (DFT) has shown to be crucial to guide experimental results in the good track \cite{Livas,9,10}. Indeed, many of the experimental discoveries had been anticipated by DFT calculations \cite{A,B,C,D,E}. Furthermore, making use of DFT-based structural prediction methods, most of the binary combinations of hydrogen and host atoms have been theoretically explored. Consequently, the most favourable features to enhance $T_c$ in electron-phonon mediated hydrogen-based superconductors has  emerged \cite{Livas, Ishikawa, Semenok, 25}: weakened covalent bonds, regular atomic bonding patterns, and high density of states at the Fermi level. So far, however, the stability of binary hydrides having a predicted superconducting critical temperature above 100 K has been limited to pressures above 100 GPa, with the exception of RbH$_{12}$, calculated to be stable at 50 GPa with a $T_c$ of about 115 K \cite{11}.

The focus is now shifting towards reducing the pressure of stability for these hydrogen-rich systems. To further expand the list of predicted compounds, in recent years attempts have been made to explore the energy landscape of ternary hydrides \cite{I,II,III,IV,V}. Two of the two most prominent results are the prediction of a $T_c$ of about 450 K at 250 GPa for MgLi$_2$H$_{16}$ \cite{III} and the possible metastability down to 40 GPa of LaBH$_8$ in the $Fm\bar{3}m$ high-symmetry phase with a remarkable $T_c$ of approximately 120 K \cite{13,14}. This latter result suggests that with the right ternary combination of atoms it is possible to synthesize phases at high pressures that could remain metastable down to room pressure. The same structural motifs of LaBH$_8$ with a similar critical temperature have also been recently predicted for BaSiH$_8$ and SrSiH$_8$ \cite{Lucrezi}. This further confirms that high-$T_c$ ternary hydrogen-rich compounds may be (meta)stable at ambient pressure.

In most of the DFT-based calculations for hydrogen-based superconductors at high pressure, however, the ions in the system are treated as classical particles. This means they are considered as a fixed point at the local minima of the $V(\mathbf{R})$ Born-Oppenheimer energy surface and the vibrational phonon frequencies are determined from the second derivatives of $V(\mathbf{R})$ taken at the minimum. The vector $\mathbf{R}$ represents the position of all atoms in the crystal. Nevertheless, for this specific class of systems composed mainly of light hydrogen atoms, it is important to have in mind that quantum ionic fluctuations cannot be neglected and can significantly alter the structural, phononic, and superconducting properties. For instance, the crystal structure for which the highest critical temperature has been observed in H$_3$S and LaH$_{10}$ is only stable thanks to quantum effects \cite{15,Errea2015High,16}, the structural properties of the predicted superconducting ScH$_6$ are also largely affected \cite{BB}, and the lattice anharmonicity that arises due to quantum lattice fluctuations introduces a large renormalization in the phonon spectra of PdH \cite{CC} and AlH$_3$ \cite{AA}. It seems that, as illustrated by the H$_3$S and LaH$_{10}$ cases, quantum effects tend to stabilize structures with a large electron-phonon coupling and keep them stable at much lower pressures than expected with standard calculations treating the ions classically.

In this work we aim at investigating the effect of quantum ionic fluctuations in LaBH$_8$ through \emph{ab initio} DFT calculations and the use of the stochastic self-consistent harmonic approximation (SSCHA) \cite{ErreaSSCHA,BiancoSSCHA,Monacelli2018Pressure,MonacelliSSCHA}. We study the thermodynamic stability and superconducting properties of this compounds with the hope that quantum effects could make LaBH$_8$ metastable even at ambient pressure. The obtained results show that quantum effects and the consequent anharmonicity enhance the superconducting critical temperature, but, contrary to the expectations, tend to destabilize the $Fm\bar{3}m$ crystal structure, making it dynamically unstable below approximately 77 GPa.

\section{Theoretical methods and calculation details}

\subsection{The stochastic self-consistent harmonic approximation}

The SSCHA \cite{ErreaSSCHA,BiancoSSCHA,Monacelli2018Pressure,MonacelliSSCHA} is a stochastic method based on a variational minimization of the free energy that includes the effect of quantum ionic fluctuations and does not approximate the $V(\mathbf{R})$ Born-Oppenheimer potential, keeping all the anharmonic terms in it. The SSCHA minimizes the free energy calculated with a trial density matrix $\widetilde\rho_{\mathbfcal{R},\mathbf{\Phi}}$ as
\begin{equation}
    \mathcal{F}[\widetilde \rho_{\mathbfcal{R},\mathbf{\Phi}}] = \mathrm{tr}[\widetilde \rho_{\mathbfcal{R},\mathbf{\Phi}} H] + \frac{1}{\beta} \mathrm{tr}[\widetilde \rho_{\mathbfcal{R},\mathbf{\Phi}} \ln( \widetilde \rho_{\mathbfcal{R},\mathbf{\Phi}})],
\end{equation}
where $\beta=1/(k_BT)$, with $T$ the temperature and $k_B$ Boltzmann's constant, and $H = K + V(\textbf{R})$ is the full ionic Hamiltonian, including both the kinetic energy $K$ and the full $V(\textbf{R})$ potential. The trial density matrix $\widetilde \rho_{\mathbfcal{R},\mathbf{\Phi}}$ is chosen so that the probability distribution of ionic positions that it defines, i.e. $\widetilde \rho_{\mathbfcal{R},\mathbf{\Phi}}(\mathbf{R})$, is a multidimensional Gaussian centered at the centroid positions $\mathbfcal{R}$, whose width is related to the matrix $\mathbf{\Phi}$ (bold symbols represent vectors, matrices, and tensors in compact notation). The SSCHA minimizes $\mathcal{F}[\widetilde \rho_{\mathbfcal{R},\mathbf{\Phi}}]$ as a function of $\mathbfcal{R}$ and $\mathbf{\Phi}$. At the minimum, the obtained $\mathbfcal{R}_{eq}$ positions determine the average ionic positions and the auxiliary force constants $\mathbf{\Phi}_{eq}$ are related to the fluctuations around these positions. In contrast, in the classical harmonic approximation the positions are fixed by the $\mathbf{R}_0$ positions that minimize $V(\mathbf{R})$, which, in general, are different from $\mathbfcal{R}_{eq}$ because the ionic kinetic energy is only accounted for by the latter. 

In the static limit of the SSCHA theory \cite{BiancoSSCHA,MonacelliSSCHA,Monacelli2021Time,Lihm2021Gaussian}, phonon frequencies are determined from the eigenvalues of the Hessian of the free energy taken at $\mathbfcal{R}_{eq}$,
\begin{equation}
\mathcal{D}_{ab}^{(F)} = \frac{1}{\sqrt{M_aM_b}} \frac{\partial^2 \mathcal{F}}{\partial\mathcal{R}_a\partial\mathcal{R}_b}\bigg|_{\mathbfcal{R}_{eq}},
 \end{equation}
where $a$ labels both an atom and a Cartesian index and $M_a$ is the mass of atom $a$. The dynamical matrix $\mathbfcal{D}^{(F)}$ defined from the free energy Hessian is the quantum anharmonic analogue to the classical harmonic dynamical matrix, which is defined instead from the Hessian of the Born-Oppenheimer potential taken at $\mathbf{R}_0$: 
\begin{equation}
\mathcal{D}_{ab}^{(h)} = \frac{1}{\sqrt{M_aM_b}} \frac{\partial^2 V(\mathbf{R})}{\partial R_a\partial R_b}\bigg|_{\mathbfcal{R}_{0}}.
 \end{equation}
It is worth noting that a negative eigenvalue of $\mathbfcal{D}^{(F)}$ signals that the structure is unstable in the free energy landscape that includes quantum effects and anharmonicity. In the same way, a negative eigenvalue of $\mathbfcal{D}^{(h)}$ signals a classical instability when ionic quantum effects are neglected. The quantum anharmonic free energy dynamical matrix has an explicit expression in the SSCHA \cite{BiancoSSCHA}, 
\begin{equation}
    \mathbfcal{D}^{(F)} = \mathbfcal{D}_{eq}^{(2)} + \mathbfcal{D}_{eq}^{(3)}: \mathbf{\Lambda}_{eq}:\bigg[ \mathbb{1}  - \mathbfcal{D}_{eq}^{(4)}:\mathbf{\Lambda}_{eq} \bigg]^{-1} : \mathbfcal{D}_{eq}^{(3)}
    \label{eq:Df}
\end{equation}
where $\mathcal{D}_{ab,eq}^{(2)}=\Phi_{ab,eq}/\sqrt{M_aM_b}$, $\mathbf{\Lambda}_{eq}$ is a fourth order tensor that depends on the eigenvalues and eigenvectors of $\mathbfcal{D}_{eq}^{(2)}$ \cite{BiancoSSCHA}, and the non-perturbative higher-order force constants are calculated as
\begin{equation}
    \mathbfcal{D}_{a_1 \cdots a_n}^{(n)} = \frac{1}{\sqrt{M_{a_1} \cdots M_{a_n}}}  \bigg\langle\frac{\partial^nV(\textbf{R})}{\partial R_{a_1} \cdots \partial R_{a_n}} \bigg\rangle_{\widetilde \rho_{\mathbfcal{R}_{eq},\mathbf{\Phi}_{eq}}}.
    \label{eq:Dn}
\end{equation}
In Eq. \eqref{eq:Df} the double dot denotes a tensor product and in Eq. \eqref{eq:Dn} the $\langle\rangle_{\widetilde \rho_{\mathbfcal{R}_{eq},\mathbf{\Phi}_{eq}}}$ denotes the quantum statistical average taken with the density matrix at the end of the minimization (see Ref. \onlinecite{MonacelliSSCHA} for more details about the notation).

The SSCHA, apart from optimizing the internal ionic positions through the minimization of $\mathcal{F}[\widetilde \rho_{\mathbfcal{R},\mathbf{\Phi}}]$ with respect to the centroid positions, can also relax the lattice cell parameters including quantum effects and anharmonicity \cite{Monacelli2018Pressure}. This is obtained by calculating the stress tensor from the derivative of the SSCHA free energy with respect to the components of a strain tensor $\boldsymbol{\epsilon}$: 
 \begin{equation}
     P_{\alpha\beta} = - \frac{1}{\Omega_{Vol}} \frac{\mathcal{\partial F}}{\partial \epsilon_{\alpha\beta} } \bigg|_{\epsilon=0},
     \label{Eq:Pressure}
\end{equation}
where $\Omega_{Vol}$ is the volume of the cell and $\alpha$ and $\beta$ are Cartesian indexes. This expression is able to take into account the additional pressure induced by ionic fluctuations on top of the classical harmonic pressure, which is calculated by replacing in Eq. \eqref{Eq:Pressure} the SSCHA free energy with $V(\textbf{R})$.

\subsection{The electron-phonon interaction}

We calculate the Eliashberg function as
\begin{equation}
    \alpha^2F(\omega) = \frac{1}{2\pi N(\epsilon_F) N_q} \sum_{\mu \mathbf{q}} \frac{\gamma_\mu(\mathbf{q})}{\omega_\mu(\mathbf{q})}\delta(\omega-\omega_\mu(\mathbf{q})),
 \label{eq:a2f}
 \end{equation}
 where
  \begin{multline}
\gamma_\mu(\mathbf{q}) = \frac{\pi}{N_k} \sum_{\mathbf{k}nm} \sum_{ab}\frac{\epsilon_\mu^a(\mathbf{q}) \epsilon_\mu^b(\mathbf{q})^*}{\sqrt{M_aM_b}} d^a_{\mathbf{k}n,\mathbf{k+q}m} d^{b*}_{\mathbf{k}n,\mathbf{k+q}m} \\
\times \delta(\varepsilon_{\mathbf{k}n}-\varepsilon_F)\delta(\varepsilon_{\mathbf{k+q}m}-\varepsilon_F)
\label{eq:linewidth}
 \end{multline}
is the phonon linewidth associated to the electron-pohonon interaction of the phonon mode $\mu$ at wavevector $\mathbf{q}$. In Eqs. \ref{eq:a2f} and \ref{eq:linewidth} $N_k$ and $N_q$ are the number of electron and phonon momentum points used for the Brillouin zone sampling, $N(\epsilon_F)$ is the density of states (DOS) at the Fermi level, $\omega_\mu(\mathbf{q})$ and $\epsilon_{\mu}^a(\mathbf{q})$ represent phonon frequencies and polarization vectors, and 
\begin{equation}
d^a_{\mathbf{k}n,\mathbf{k+q}m} = \bra{\mathbf{k}n}\delta V_{KS}/\delta R^a(\mathbf{q})\ket{\mathbf{k+q}m}
\end{equation}
is the electron-phonon matrix element between Kohm-Sham states  $\ket{\mathbf{k}n}$ (with $n$ a band index and $\bf k$ a wavevector index) and $\ket{\mathbf{k+q}m}$ of the derivative of the Kohm-Sham potential $V_{KS}$ with respect to the Fourier transformed displacement of the $a$th atom. $\varepsilon_{\mathbf{k}n}$ is the energy of the $\ket{\mathbf{k}n}$ state and $\varepsilon_F$ is the Fermi energy. The electron-phonon coupling constant $\lambda$ and the average logarithmic phonon frequency can be obtained directly from the Eliashberg function as
\begin{equation}
    \lambda =  2 \int_0^{\infty} d\omega \frac{\alpha^2F(\omega)}{\omega}
\end{equation}
 and
 \begin{equation}
     \omega_{log} = \exp \bigg( \frac{2}{\lambda} \int_0^{\infty} d\omega\frac{\alpha^2F(\omega)}{\omega} log(\omega)\bigg).
 \end{equation}
Isotropic Migdal-Eliashberg \cite{Migdal} equations can also be directly solved once $\alpha^2F(\omega)$ is known.  
 
\subsection{Calculation Details}

\begin{table}[t!]
\caption{\label{Table:Wyckoff} Values of the hydrogen Wyckoff parameter for the classical structure ($x_c$), the shortest hydrogen-boron distance ($d^{H-B}_c$), and the related classical pressure ($P_c$) at the lattice parameter $a$. For the same lattice parameters we report the hydrogen Wyckoff parameter ($x_q$), the shortest hydrogen-boron distance ($d^{H-B}_q$), and the associated pressure obtained through Eq. \ref{Eq:Pressure} ($P_q$) calculated after the quantum SSCHA structural relaxation.}
\begin{tabular}{c|ccc|ccc}
  $a$(\AA)         & $x_c$   & $d^{H-B}_c$(\AA)  & $P_c$(GPa)  & $x_q$   & $d^{H-B}_q$(\AA)  & $P_q$(GPa) \\ 
\hline 
  5.577               & 0.1459  & 1.409              & 50           & 0.1489  & 1.438              & 58          \\
  5.427               & 0.1474  & 1.386              & 75           & 0.1498  & 1.408              & 84          \\
  5.311               & 0.1485  & 1.366              & 100          & 0.1506  & 1.385              & 110         \\
\end{tabular}
\end{table}

The DFT calculations were performed with the plane-wave {\sc Quantum ESPRESSO}\cite{18,19} package using ultrasoft pseudo-potentials with the Perdew-Burke-Ernzerhof \cite{Perdew} parametrization of the exchange-correlation potential. We included 3 electrons of boron and 11 of lanthanum in the valence.  The cutoff for the wavefunctions and density were chosen as 90 Ry and 900 Ry, respectively. The integration over the Brillouin zone in the self-consistent calculations was performed with a first-order Methfessel-Paxton smearing of 0.02 Ry broadening and a 30$\times$30$\times$30 $\mathbf{k}$-point grid. The electronic properties such as electron localization function (ELF) \cite{20,21,22} and Bader \cite{23} charge were calculated using the {\sc Quantum ESPRESSO} post processing tools.

The harmonic phonon calculations were performed on a 6$\times$6$\times$6 phonon $\bf q$-point grid making use of density functional perturbation theory (DFPT) \cite{Baroni2001Phonons}. The SSCHA calculations were performed on a 2$\times$2$\times$2 supercell containing 80 atoms, which yielded $\mathbfcal{D}^{(F)}(\bf q)$ anharmonic dynamical matrices in a commensurate 2$\times$2$\times$2 $\bf q$-point grid. The difference between the harmonic and anharmonic dynamical matrices was Fourier interpolated to the finer 6$\times$6$\times$6 grid, and, by adding to the interpolated result the harmonic dynamical matrices, $\mathbfcal{D}^{(F)}(\bf q)$ was obtained in the  6$\times$6$\times$6 grid. The SSCHA minimization was converged using a 7$\times$7$\times$7 coarser $\mathbf{k}$-point grid for the DFT self-consistent calculations in the supercells needed to get the forces for the stochastic minimization of the free energy. The SSCHA was run at 0 K. 

The electron-phonon interaction was calculated both in the classical harmonic and quantum anharmonic cases. In both cases a 40$\times$40$\times$40 $\mathbf{k}$-point grid was used in Eq. \eqref{eq:linewidth}, with a Gaussian smearing of 0.008 Ry to approximate the Dirac deltas in the equation, and 6$\times$6$\times$6 $\bf q$-point grid. For the classical harmonic case the electron-phonon interaction was calculated having the atoms sitting at the positions $\mathbf{R}_0$ that minimize $V(\mathbf{R})$, while in the anharmonic case the atoms were placed at the positions $\mathbfcal{R}_{eq}$ obtained through the SSCHA minimization. 

The superconducting critical temperature was calculated by solving the isotropic Migdal-Eliashberg equations \cite{Migdal} with values of $\mu$ between 0.1 and 0.15, with 20 to 30 Matsubara frequencies, and an energy cutoff of about 50 eV. The Eliashberg function is calculated both at the harmonic or anharmonic levels, respectively, by plugging into Eqs. \ref{eq:a2f} and \ref{eq:linewidth} the  harmonic  phonon  frequencies  and polarization vectors or their anharmonic counterparts obtained diagonalizing $\mathbfcal{D}^{(F)}$.


\section{Results and discussion}

\subsection{Structural and electronic properties}

\begin{figure}
\vspace{6mm}
\includegraphics[scale=0.35]{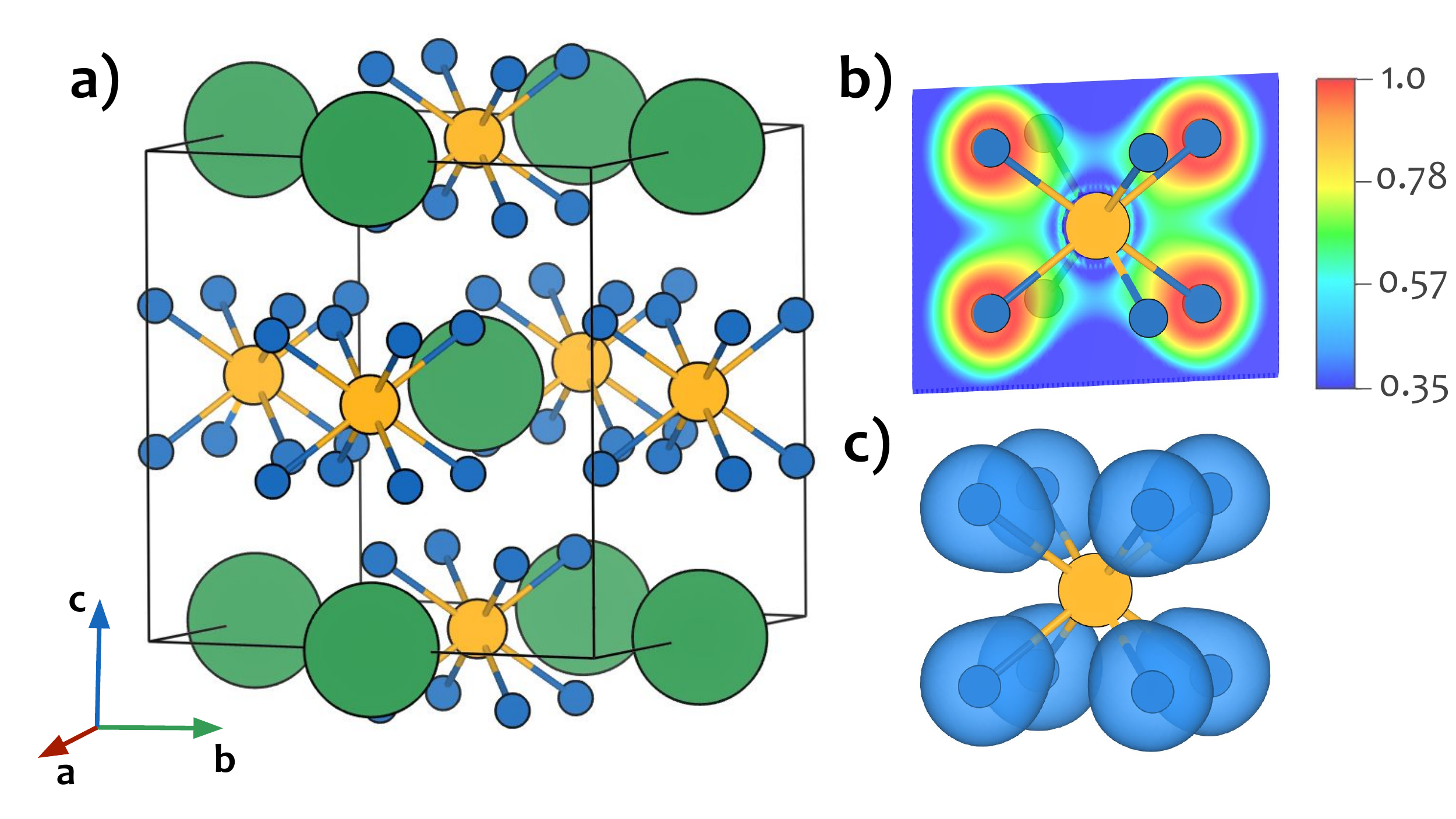}
\caption{\label{Fig:Structure_ELF} (a) Conventional unit cell of $Fm\bar{3}m$ LaBH$_8$. (b) Cross section of the ELF for the BH$_8$ unit between values of 1 and 0.35. (c) The ELF isosurface of the BH$_8$ unit at a value of 0.69.}
\end{figure}

The face-centered cubic unit cell of the superconducting LaBH$_8$ shown in Fig. \ref{Fig:Structure_ELF} respects a $Fm\bar{3}m$ symmetry, with the boron, lanthanum, and hydrogen atoms located, respectively, at the 4a, 4b, and 32f Wyckoff sites. While the La and B 4a and 4b sites are completely fixed by symmetry, the H 32f site has a free parameter. A representative of the 32f orbit  can be written as $(x,x,x)$. Thus, the entire structure can be determined by only two free parameters: the lattice parameter $a$ and the $x$ parameter related to the hydrogen 32f sites.
\begin{figure}
\includegraphics[scale=0.33]{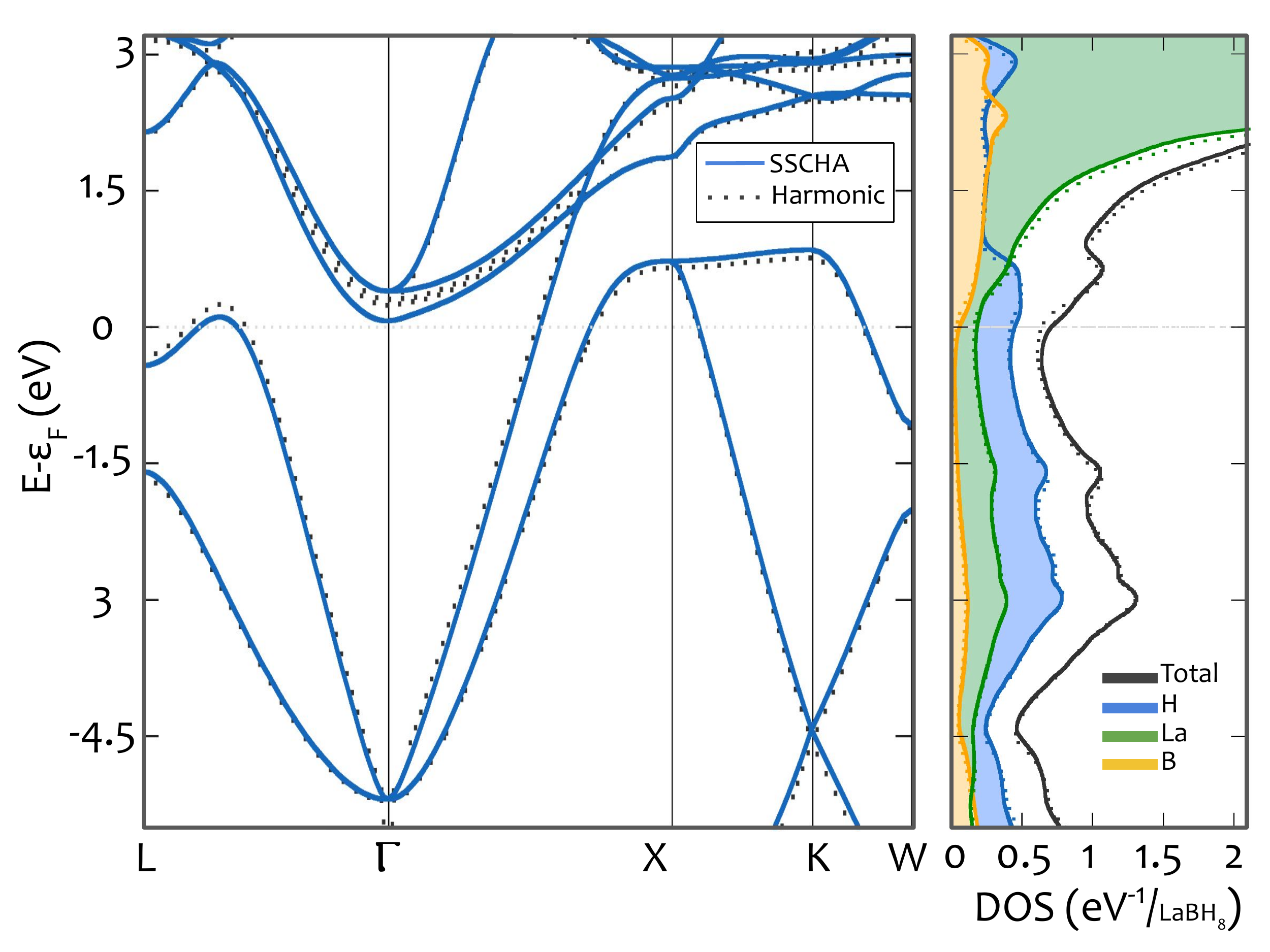} %
\caption{\label{Fig:EBANDS} (Left panel) Electronic band structure with the atoms at the $\mathbf{R}_0$ classical harmonic sites and at the quantum anahrmonic $\mathbfcal{R}_{eq}$ sites after the SSCHA relaxation, in both cases with $a =5.577$ {\AA}. (Right panel) Total DOS and its projections onto different atoms calculated both at the classical harmonic positions (dotted lines) and quantum SSCHA positions (solid lines).}
\end{figure}

A classical structural relaxation of the system, in which the the $x$ and $a$ parameters are determined from $V(\bf R)$, was performed at 50, 75, and 100 GPa. After the relaxation, the lattice parameter and the hydrogen Wyckoff parameter were found to be, respectively, 5.577 {\AA} and 0.14592 at 50 GPa, 5.427 {\AA} and 0.1474 at 75 GPa and 5.311 {\AA} and 0.1485 at 100 GPa. The boron-hydrogen shortest distance decreases with increasing pressure, passing from a value of 1.409 {\AA} at 50 GPa to 1.366 {\AA} at 100 GPa. The structural relaxation was then repeated using the SSCHA for the same lattice parameters obtained for the classical results. After the SSCHA relaxation, the $x$ hydrogen coordinates changed from 0.1459 to 0.1489 at $a = 5.577$ {\AA} (classical pressure of 50 GPa), from 0.1474 to 0.1498 at $a = 5.427$ {\AA} (classical pressure of 75 GPa), and from 0.1485 to 0.1506 at $a = 5.311$ {\AA} (classical pressure of 100 GPa). Using Eq. \eqref{Eq:Pressure} to compute the quantum pressure, it was found that quantum effects add respectively  8 GPa, 9 GPa, and 10 GPa on top of the classical 50 GPa, 75 GPa, and 100 GPa pressures, respectively, which is in line with the pressure exerted by ionic quantum fluctuations in other hydrogen-rich compounds \cite{15,16}. Table \ref{Table:Wyckoff} summarizes the differences between the quantum and classical pressures and the $x$ parameters for the classical and SSCHA structural relaxations for different lattice parameters. Overall, the introduction of quantum effects through the SSCHA generates an increment in pressure and an expansion of the boron-hydrogen distance for the same lattice parameter of about 2\%. This behavior is attributed to the extra space that the quantum atoms occupy with respect to their classical counterparts.

\begin{figure*}[t!]
\includegraphics[scale=0.7]{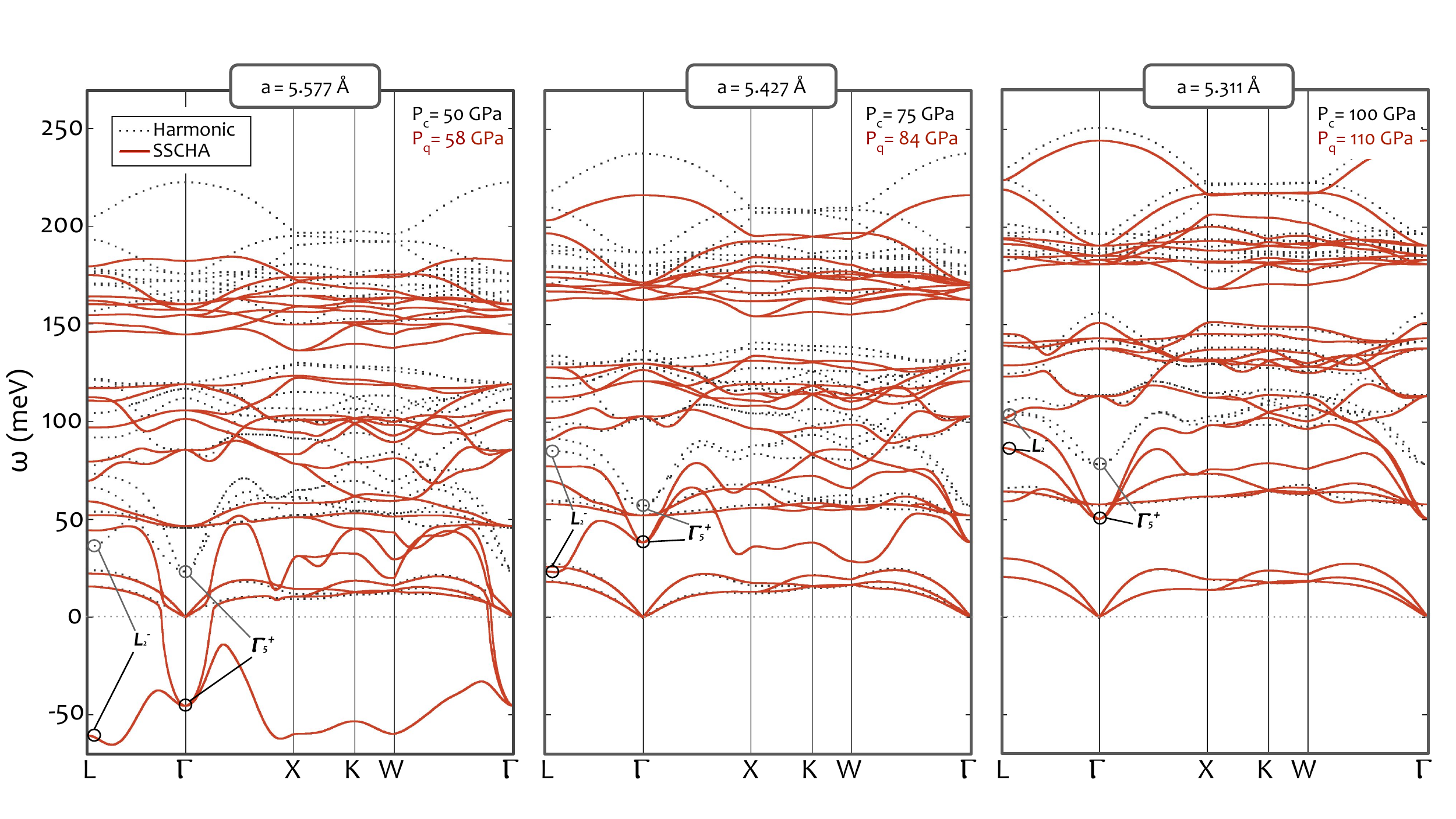}%
\caption{\label{Fig:PBANDS} Phonon spectra calculated for the lattice parameters reported in Table \ref{Table:Wyckoff} at the harmonic and anharmonic (SSCHA) level. The harmonic phonon energies are calculated from $\mathbfcal{D}^{(h)}$ with the hydrogen atoms at the $(x_c,x_c,x_c)$ sites. The SSCHA anharmonic phonons are calculated from $\mathbfcal{D}^{(F)}$ with the hydrogen atoms at the $(x_q,x_q,x_q)$ sites. The introduction of the quantum effects increases the pressure for the same lattice parameter.}
\end{figure*}

An analysis of the electronic band structure is shown in Fig. \ref{Fig:EBANDS}. The calculations were performed for both classical and SSCHA  structures corresponding to a lattice parameter of 5.577 {\AA}, which corresponds to a classical pressure of $P_c= 50$ GPa and a quantum pressure of $P_q= 58$ GPa. The results show that the variation of the atomic Wyckoff parameter introduces just small variations of the order of 150 meV in the band structure around the Fermi energy. However, these variations are not significant enough to alter the overall shape of the bands. With respect to the DOS at the Fermi level, quantum effects produce a slight increment of its value. The highest contribution to the DOS comes from hydrogen and lanthanum atoms, which  contribute respectively a 67.5\% and a 26.4\% to the total DOS at the Fermi level in the classical case, values that are slightly adjusted to 65\% and 26.4\% with the inclusion of ionic quantum effects.

The analysis of the electron localization function (ELF) shows how the LaBH$_8$ system is composed by BH$_8$ units, where the boron atom is covalently bonded with eight hydrogen atoms as shown in Fig. \ref{Fig:Structure_ELF}. The appearance of the covalent bonds is identified by an elongation of the ELF isosurfaces around the hydrogen atoms towards the boron atom \cite{22}. Furthermore, the analysis of the Bader charge shows the La atom donates 1.49 electrons to the BH$_8$ unit, a feature that suggests the existence of an ionic bond between the La atoms and the BH$_8$ units. 

Finally, the behavior of the networking value\cite{25} ($\phi$) was analyzed for the classical and quantum structures for $a=5.577$ {\AA} ($P_c = 50$ GPa) and $a=5.311$ {\AA} ($P_c= 100$ GPa). The networking value is the highest value of the ELF that creates an isosurface that expands periodically in all the crystal. It has been recently shown that $\phi$ correlates well with $T_c$ \cite{25}, implying that the larger the networking value, the greater the expected $T_c$. For the classical and quantum structures at  $a=5.577$ {\AA}, the respective values of $\phi$ are 0.28 and 0.30. The network is constructed by cores of BH$_8$ units, as shown in Fig. \ref{Fig:Structure_ELF}b, that weakly bond with each other. According to the ELF classification in Ref. \onlinecite{25}, the bonding between adjacent BH$_8$ units appears at values of ELF too low to be considered as a weak covalent hydrogen-hydrogen system, type of bonding that yields the highest $T_c$'s. Combined with the fraction of hydrogen atoms in the stoichiometry ($H_f$) and the hydrogen fraction of the DOS at the Fermi energy ($H_{\mathrm{DOS}}$), the networking value gives rise to an expected value for the critical temperature of 60 K and 70 K for the classical and quantum case, respectively, according to the $T_c=(750 \phi H_f H_{\mathrm{DOS}}^{1/3} - 85)$ K formula \cite{25}. At the smaller lattice parameter $a=5.311$ {\AA} for the classical and quantum structure we obtained respectively values of $\phi=0.30$ and $\phi=0.32$, predicting critical temperatures of 70 K and 80 K, respectively. The networking value analysis suggests, thus, that quantum effects will enhance the critical temperature in $Fm\bar{3}m$ LaBH$_8$.

\subsection{Phonon properties and lattice instabilities}

\begin{figure}
\includegraphics[scale=0.35]{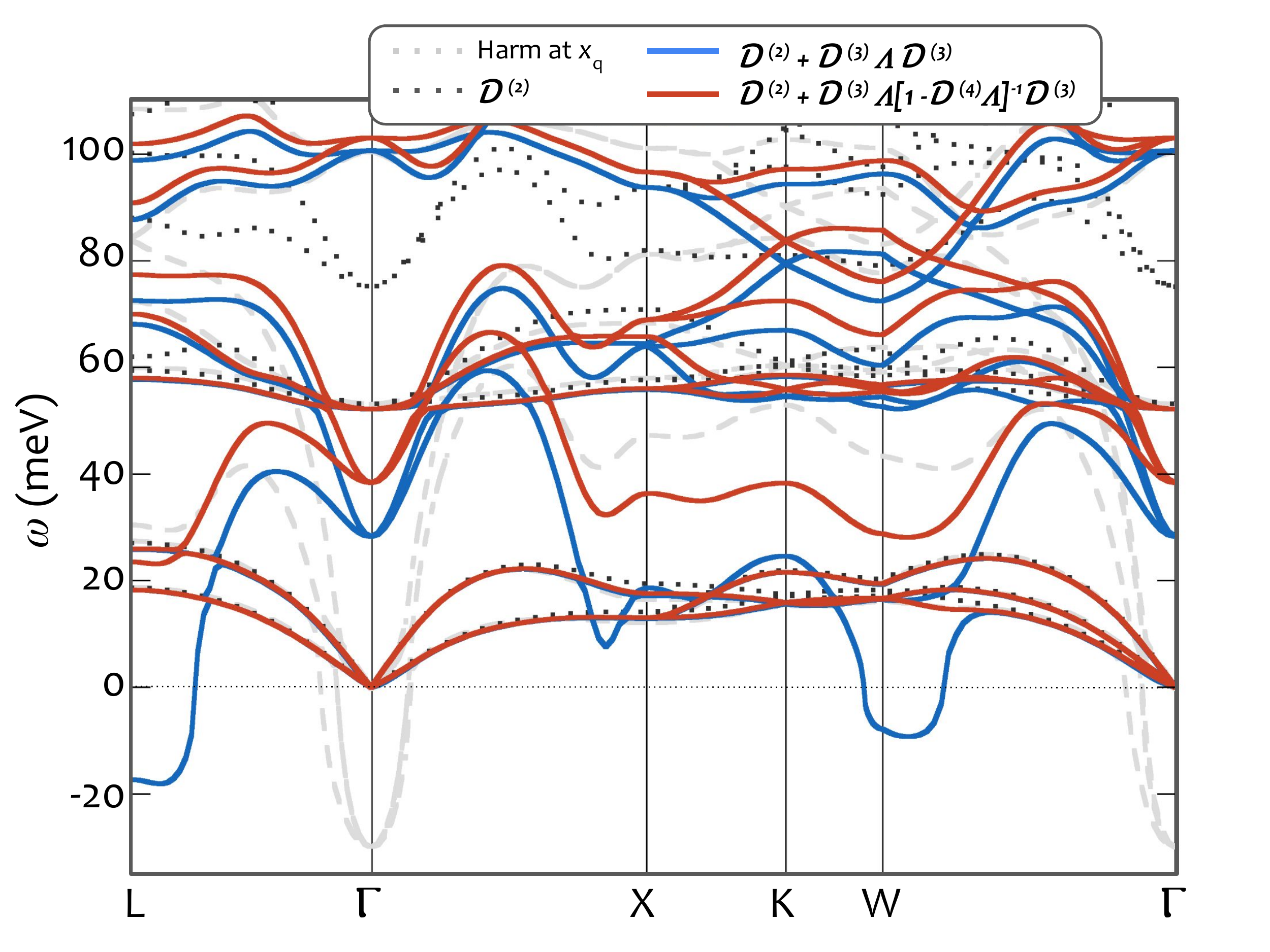}
\caption{\label{Fig:Order} Harmonic phonons with $a=5.427$ \AA\ with the H atoms at the quantum positions $(x_q,x_q,x_q)$, together with the different contributions to the SSCHA free energy Hessian $\mathbfcal{D}^{(F)}$. The figure shows the spectra only at low energies to illustrate better the effect of the different terms.}
\end{figure}

\begin{figure*}
\includegraphics[scale=0.7]{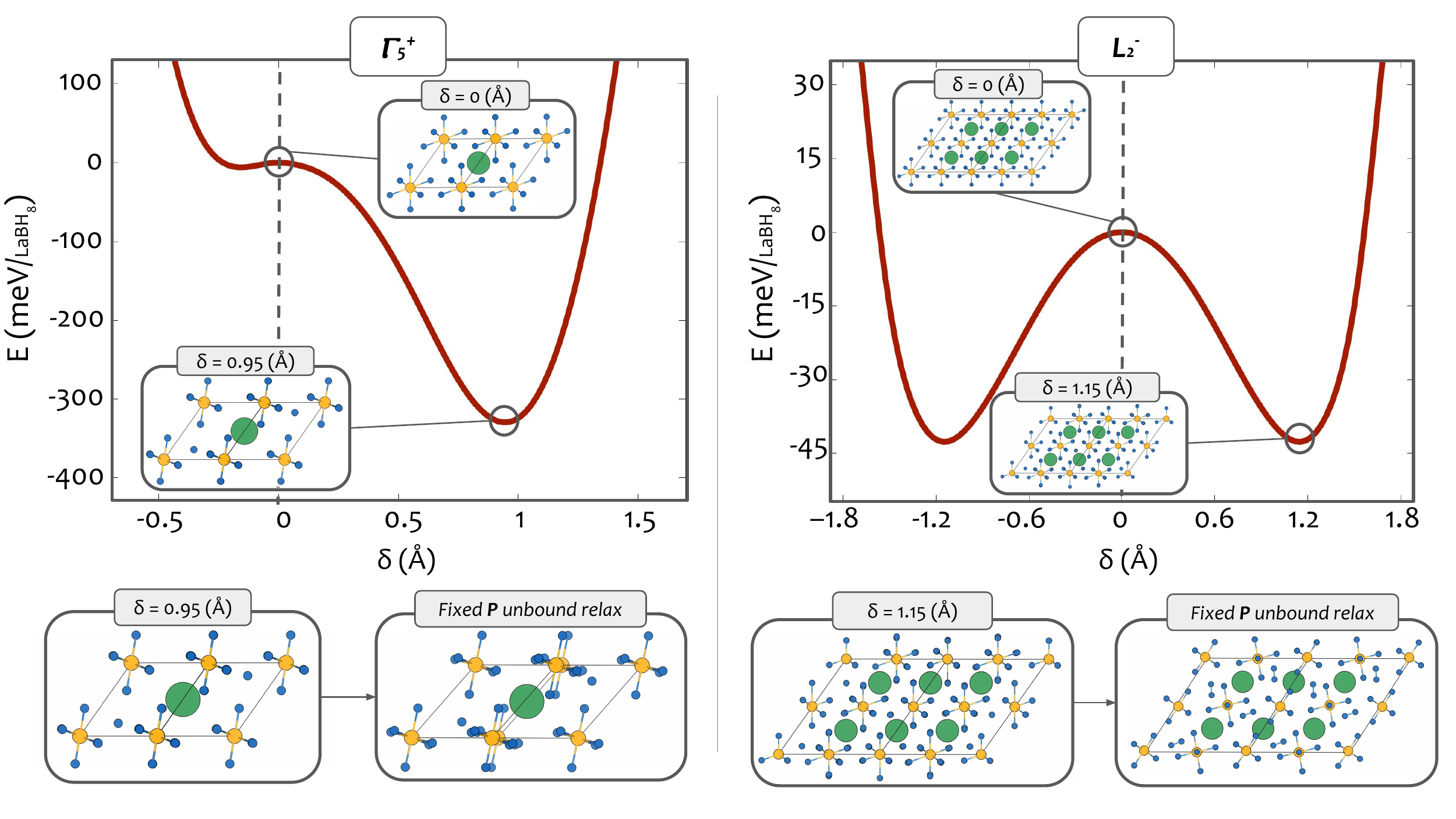}
\caption{\label{Fig:Instability} Born-Oppenheimer energy variation occurring along the distortion patterns introduced by the $\Gamma_5^+$ and the $L_2^-$ modes starting with the H atoms at the $(x_q,x_q,x_q)$ Wyckoff positions obtained after the SSCHA structural relaxation at $a = 5.577$ {\AA}. The LaBH$_8$ unit cells before the distortion, at the minimum of the potential energy curve, and after a classical cell relaxation at fixed $P_c = 50$ GPa are also shown in each case.}
\end{figure*}

The phonon spectra calculated at the classical harmonic level with the atoms at the $\mathbf{R}_0$ positions and at the quantum anharmonic level with atoms at the $\mathbfcal{R}_{eq}$ positions are shown in Fig. \ref{Fig:PBANDS}. The calculations at the highest pressures with $a=5.311$ \AA\ show that the spectrum is characterized by three distinctive regions separated with energy gaps. The acoustic modes have a predominant La character. The group of phonon modes in the 50-150 meV energy range have both boron and hydrogen character. The bands with dominant B behavior show a very weak dispersion, approximately at 57 meV. The other modes in this range are dominated by hydrogen bond bending modes. Higher energy bands, above $\sim$ 175 meV, mostly involve hydrogen bond stretching vibrations. At larger lattice parameters, i.e. lower pressure, the H bond-bending modes soften, and a mode with irreducible representation $\Gamma_5^+$, also labeled as $T_{2g}$, becomes unstable below $P_c=45$ GPa in the harmonic approximation.

\begin{figure*}[t!]
\includegraphics[scale=0.7]{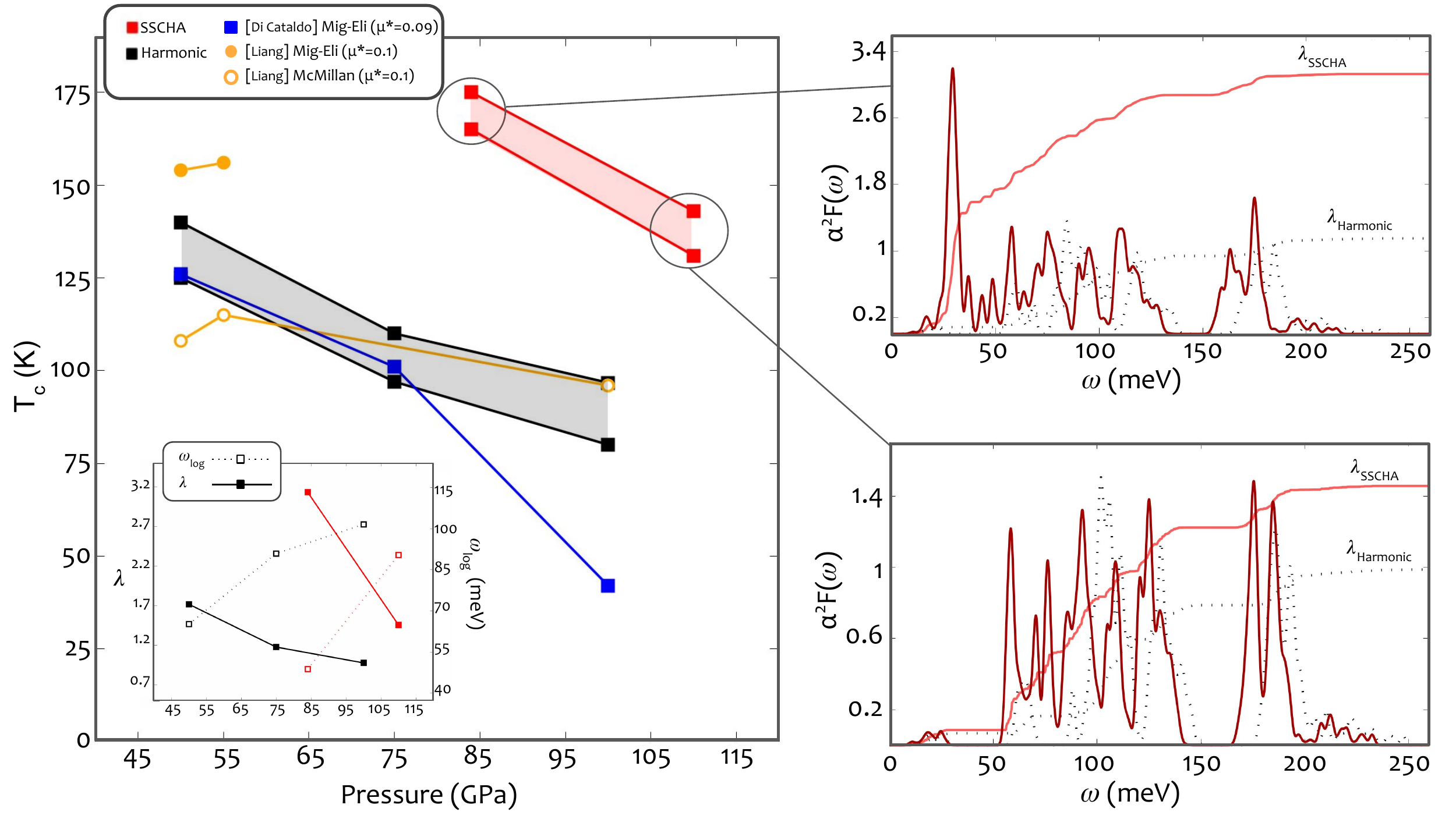}
\caption{\label{Fig:Tc} Superconducting critical temperatures obtained solving the isotropic Migdal-Eliashberg equations with values of $\mu^*$ of 0.1 and 0.15 at the classical harmonic level and quantum anharmonic level, the latter obtained with the SSCHA method. The pressure reported in the graph for the classical result is the classical pressure $P_c$, while the pressure in the SSCHA results is the quantum pressure $P_q$. The results are compared to the previous calculations by Di Cataldo \emph{et al.} \cite{13} and Liang \emph{et al.} \cite{14}. The inset shows the pressure dependence of the electron-phonon coupling $\lambda$ and $\omega_{log}$. The right panels show the $\alpha^2F(\omega)$ function together with the integrated electron-phonon coupling constant $\lambda(\omega) = 2 \int_0^{\omega} d\omega' \alpha^2F(\omega')/\omega'$. }
\end{figure*}

The quantum expansion of the BH$_8$ unit produces dramatic effects on the phonon energies, mainly introducing a large overall softening of all hydrogen-character phonon modes. La and B character modes are, on the contrary, barely affected by quantum anharmonic effects, even if hydrogen bond-bending modes get mixed with the lanthanum acoustic modes at low pressures. The main effect of quantum effects is that the structure becomes unstable at higher pressures than in the classical calculations. While the classical structure is stable for $a=5.577$ \AA\ ($P_c = 50$ GPa), $\mathbfcal{D}^{(F)}$ has large imaginary eigenvalues at different points of the Brillouin zone for the same lattice parameter ($P_q = 58$ GPa). As shown in Fig. \ref{Fig:Order}, it is remarkable that in the calculation of $\mathbfcal{D}^{(F)}$ the $\mathbfcal{D}^{(4)}$ term in Eq. \eqref{eq:Df} has a large impact and cannot be neglected, contrary to what happens in many charge-density wave materials \cite{Zhou2020Anharmonicity,Zhou2020Theory,Bianco2020Weak} and other hydrogen-based superconductors \cite{16,AA,BB}. This is specially true around the pressures at which the instabilities appear. It is also worth mentioning that, even if the harmonic phonons with $a=5.427$ \AA\ are stable if the H atoms occupy the classical $(x_c,x_c,x_c)$ sites (see Fig. \ref{Fig:PBANDS}), the structure is moved to a saddle point of the Born-Oppenheimer energy surface when the H atoms sit instead at the $(x_q,x_q,x_q)$ sites, which is illustrated by the imaginary frequencies obtained at the harmonic level for this position (see Fig. \ref{Fig:Order}). 

In our calculations including quantum anharmonic effects, $Fm\bar{3}m$ LaBH$_8$ becomes unstable at a pressure of approximately 77 GPa. The instability is led  by an optical phonon mode at the L point with $L_2^-$ symmetry, associated with a bond bending oscillation of the H atoms. The  $\Gamma_5^+$ mode, which describes a rhomboidal-like distortion of the BH$_8$ unit, becomes unstable soon after at 70 GPa. At lower pressures, the whole branch becomes unstable pointing towards a clear quantum instability of the BH$_8$ unit. In Ref. \onlinecite{13} the classical instability of the $\Gamma_5^+$ mode was estimated at 35 GPa, which is consistent with the 45 GPa pressure we obtain, and that anharmonic effects only increased this pressure by 5 GPa. In our calculations, however, quantum effects increase the destabilization pressure by 32 GPa. The difference is that in the previous calculation anharmonicity was considered neglecting its effect on the atomic positions and considering exclusively the anharmonic self-interacting terms of the $\Gamma_5^+$ mode. Our results show that anharmonicity cannot be treated in such a simplistic model, as the latter cannot describe the impact of anharmonicity on the whole phonon spectrum as well as the ionic positions.

Starting by the structure obtained after the SSCHA relaxation with the lattice parameter 5.577 {\AA}, the distortion patterns introduced by the $\Gamma_5^+$ and $L_2^-$ modes were further studied by deforming the structure according to the corresponding eigenvectors, without changing the lattice, until reaching a local energy minima as shown in Fig. \ref{Fig:Instability}. For the case of the triply degenerate $\Gamma_5^+$ mode, we show the displacement pattern that yields the lowest-energy local minimum. There are four equivalent $L$ points in the Brillouin zone  and each of them has associated a $L_2^-$ mode. We again report the $L_2^-$ mode yielding the lowest energy local minimum.  The deformation of the atomic coordinate obeys the following form:
\begin{equation}
 X_{i,p}^\alpha = X_{0i,p}^\alpha + \delta  b_{\mu i,p}^\alpha,
\end{equation}
where $X_{i,p}^\alpha$ is the atomic position of atom $i$ in the unit cell $p$ along Cartesian direction $\alpha$, $\delta$ is an arbitrary deformation parameter, and $b_{\mu i,p}^\alpha$ is a real space polarization vector of the form
\begin{equation}
b_{\mu i,p}^\alpha = \sqrt{\frac{M_0}{N_p M_i}} e^{i\mathbf{q} \cdot \mathbf{R}_p} \epsilon_{\mu i}^\alpha(\mathbf{q}).
\end{equation}
Here $N_p$ is the number of unit cells commensurate with the wave number $\mathbf{q}$, $\mathbf{R}_p$ is a unit cell lattice vector, and $M_0$ is a mass renormalization parameter obtained by having the vectors $b_{\mu i,p}^\alpha$ satisfy the condition
\begin{equation}
    \sum_{i\alpha p} M_i  (b_{\mu i,p}^{\alpha})^*b_{\mu' i,p}^{\alpha} = M_0\delta_{\mu\mu'}.
\end{equation}
 After reaching the minimum energy given by the deformation, a further classical cell relaxation with a fixed pressure of 50 GPa was performed. For both distortions it was found the BH$_8$ unit dissociates into a BH$_6$ unit. For the distortion pattern that started with the $\Gamma_5^+$ mode the two remaining hydrogen atoms sit in an interstitial site, while for the structure distorted initially with the $L_2^-$ mode they merge and form a H$_2$ molecule. Comparing the classical enthalpy between the initial $Fm\bar{3}m$ structure at $P_c = 50$ GPa and the structures obtained at the end, it was found that the structure obtained after the relaxation following $\Gamma_5^+$ is 0.8 eV/LaBH$_8$ lower in enthalpy, while the structure obtained following $L_2^-$ is 1.2 eV/LaBH$_8$ lower. This results suggests that at low pressures the high-symmetry $Fm\bar{3}m$ classical structure sits in a very shallow local minimum, which is much higher in energy than other distorted structures. Ionic quantum effects are enough to take the system out of the local minimum and to destabilize the system towards decomposition. 
 
\subsection{Superconducting properties}

Despite quantum effects destabilize $Fm\bar{3}m$ at higher pressures than expected classically, the SSCHA calculations suggest it is metastable above 77 GPa (quantum pressure). Thus, it is reasonable to investigate its superconducting properties, monitoring the impact of quantum effects and anharmonicity on them. The results for $T_c$ obtained solving isotropic Migdal-Eliashberg equations are shown in Fig. \ref{Fig:Tc}. The isotropic approximation is justified, as previous results \cite{13} show a good agreement between the critical temperature obtained with the isotropic and anisotropic Migdal-Elihasberg equations. Indeed, our results at the classical harmonic level are in rather good agreement with those reported previously \cite{13,14}. 

The introduction of the structural and phonon renormalization through the SSCHA has a strong effect on the superconducting properties, contrary to previous estimates in which a negligible impact was claimed \cite{13}. Anharmonicity increases the electron-phonon coupling and reduces the average logarithmic frequency $\omega_{log}$. The increase of $\lambda$ imposed by quantum anharmonic effects is enough to induce a large enhancement of the predicted $T_c$. Looking at $\alpha^2F(\omega$), it is possible to notice that the high value of $\lambda$ at 84 GPa in the SSCHA is mostly related to the anharmonic softening of the lower optical branches, which make the electron-phonon coupling soar due to the proximity of the system to a structural instability. In general, the overall increment of $\lambda$, and consequently $T_c$, is also due to the quantum weakening of the BH$_8$ bonds and the consequent softening of the hydrogen-character phonon modes. The reason why in previous calculations \cite{13} a weak anharmonic renormalization of $T_c$ was claimed is because the impact of anharmonicity was only estimated for the $\Gamma_5^+$ mode, exclusively considering the interaction of this mode with itself.

\section{Conclusions}

In this work we show that ionic quantum effects and the consequent anharmonicity play a crucial role on the structural and superconducting properties of the recently predicted $Fm\bar{3}m$ LaBH$_{8}$ compound. Our results show that quantum effects are able to strongly alter the behavior of the system by increasing its superconducting critical temperature, but also increasing the lowest pressure of its thermodynamic stability from the previously reported 40 GPa \cite{16} to 77 GPa, a non-negligible difference of about 40 GPa. The origin of the rise of $T_c$ as well as the lowest pressure of its dynamical stability is associated to the shallow and very anisotropic energy minimum related to distortions of the BH$_8$ unit. Although the system possesses a high symmetry, the shape of the potential $V(\mathbf{R})$ related to deformations of the BH$_8$ unit is not at all symmetric. For this reason, the introduction of ionic quantum fluctuations tends to space the atoms apart, shifting the system towards a more anharmonic zone of the  $V(\mathbf{R})$ potential. At low pressures, the energy barrier making $Fm\bar{3}m$ LaBH$_{8}$ a metastable structure in the classical $V(\mathbf{R})$ potential is shallow enough for quantum effects to push the structure out towards a lower minima, which we believe it implies dissociation of the system. These results highlight the importance of performing quantum structural relaxations in superconducting hydrogen-rich compounds, including in the calculations the the kinetic energy of the ions. Generally speaking, it is reasonable to expect that ionic quantum effects will also tend to destabilize other low pressure metastable phases composed by symmetric XH$_8$ covalent bonded units interlaced by chemical precompressors, such the recently predicted BaSiH$_8$ and SrSiH$_8$ compounds \cite{Lucrezi}.

\section*{Acknowledgements}

This research was supported by the European Research Council (ERC) under the European Union's Horizon 2020 research and innovation program (Grant Agreement No. 802533). We acknowledge PRACE for awarding us access to Joliot-Curie Rome at TGCC, France.









\def\bibsection{\section*{\refname}} 

\end{document}